\documentclass[
aps,%
12pt,%
final,%
notitlepage,%
oneside,%
onecolumn,%
nobibnotes,%
nofootinbib,%
superscriptaddress,%
noshowpacs,%
centertags]%
{revtex4}

\newcommand{\pa}{\partial}
\def\reg{\mathop{\rm reg}\nolimits}
\def\sing{\mathop{\rm sing}\nolimits}
\begin{document}
\selectlanguage{english}

\title{On applications of the Maupertuis-Jacobi correspondence \\
for Hamiltonians $F(x,|p|)$ in some 2-D stationary semiclassical problems}

\author{\firstname{S.~Yu.}~\surname{Dobrokhotov}}
\email{dobr@ipmnet.ru}
\affiliation{%
A.Ishlinskii Institute for Problems in Mechanics Russian Academy of Sciences and Moscow institute of Physics and Technology (State University)
}%
\author{\firstname{D.~S.}~\surname{Minenkov}}
\email{minenkov.ds@gmail.com}
\affiliation{%
A.Ishlinskii Institute for Problems in Mechanics Russian Academy of Sciences and Moscow institute of Physics and Technology (State University)
}%

\author{\firstname{M.}~\surname{Rouleux}}
\email{rouleux@univ-tln.fr}
\affiliation{%
Aix Marseille Universit\'e, CNRS, CPT, UMR 7332, 13288 Marseille, France, Universit\'e de Toulon, CNRS, CPT, UMR 7332, 83957 La Garde, France}%

\begin{abstract}

UDK 517.9

We make use of the Maupertuis -- Jacobi correspondence,
well known in Classical Mechanics,
to simplify 2-D asymptotic formulas based on Maslov's canonical operator,
when constructing Lagrangian manifolds
invariant with respect to phase flows for Hamiltonians of the form $F(x,|p|)$.
As examples we consider Hamiltonians coming from the Schr\"odinger equation, the 2-D Dirac equation for graphene and linear water wave theory.
\end{abstract}

\maketitle

\section{Introduction}
Maupertuis -- Jacobi correspondence \cite{Arnold,Ab,Tsi} allows to relate two Hamiltonians ${\cal H}(x,p,E)$
and $H(x,p,E)$ having in common a regular energy surface $\Sigma$;
it preserves the integral curves on $\Sigma$ up to a reparametrization of time.
As it was shown in \cite{DR1,DR2} this principle is also useful in determining the semiclassical spectral asymptotics
for a selfadjoint $h$-pseudodifferential operator ${\cal H}(x,hD_x,E;h)$
having ${\cal H}(x,p,E)$ as principal semi-classical symbol.
The other Hamiltonian ${H}(x,p,E)$ is assumed to enjoy nice properties, such as local integrability near $\Sigma$.
Then we can construct some compact Lagrangian manifolds invariant by the flow of ${\cal H}(x,p,E)$,
and then determine quasimodes for ${\cal H}(x,hD_x,E;h)$
microlocalized in a neighborhood of $\Sigma$.

In this communication we want to show that Maupertuis -- Jacobi correspondence allows
to construct non compact Lagrangian manifolds
appearing in the scattering problem for ${\cal H}(x,hD_x,E;h)$,
or the problem about Green function asymptotics.
Here it is assumed that ${H}(x,p,E)$ is a Finsler symbol,
which implies the existence of special coordinates near the singular part of the Lagrangian manifold.
When combining Maupertuis -- Jacobi correspondence
with new formulas for Maslov canonical operator \cite{DMNT},
we show that the corresponding asymptotics can be presented in a rather explicit and simple form.
We restrict ourselves to 2-D case
and apply our considerations to examples from
the Schr\"odinger (or Helmholtz) equation, the two-dimansional Dirac equation for graphene and
Pseudodifferential operators from the linear water wave theory.

\section{Lagrangian manifolds invariant with respect \protect\\
to Hamiltonians $F(x,|p|)$, the eikonal coordinates
and Maupertuis -- Jacobi correspondence}

Let $F(x,z), \quad x\in \mathbb{R}^2,\quad z\in [0,\infty)$ be a smooth function, and $E$ a real parameter.
Assume that the equation $F(x,z)=E$ has the unique solution $z=1/C(x,E)$,
where $C(x,E)$ is a smooth positive bounded function, such that $C(x,E)\geq c_0(E) > 0$.
Also we assume that $|\frac{\pa F}{\pa z}(x,\frac{1}{C(x,E)})|\geq c_1(E) > 0$, where $c_0(E),c_1(E)$ are some positive constants.
Consider the Hamiltonians  $\mathcal{H}(x,p,E)=F(x,|p|)-E, \quad H(x,p,E)=C(x,E)|p|-1$ in the phase space $\mathbb R_{p,x}^4 = T^*\mathbb{R}^2$
together with the Hamiltonian systems
\begin{gather}\label{maup1}
(a)\quad  \frac{d p}{d t}=-\mathcal{H}_x,\quad \frac{d x}{d t}=\mathcal{H}_p;\qquad
(b)\quad \frac{d p}{d \tau}=-H_x,\quad \frac{d x}{d \tau}=H_p
\end{gather}
We recall that $C(x,E)|p|$ defines a (reversible) Finsler symbol on $T^*\mathbb{R}^2$  \cite{Katok,Taylor}.

Let ${\cal Q}=\mathbb{R}$ or ${\cal Q}=\mathbb{R}/2\pi\mathbb{Z}$ and 
${\cal Q}\to T^*\mathbb{R}^2$, $\varphi\mapsto(P^0(\phi,E), X^0(\phi,E))$ be a smooth embedding with image $\Lambda^1$ such that
\begin{equation*}
\mathcal{H}(X^0(\phi,E),P^0(\phi,E),E)=0 \qquad\text{and}\qquad |P^0(\phi,E)|C(X^0(\phi,E))=1
\end{equation*}
Consider the solutions $(\mathcal{P}(t,\phi,E),\mathcal{X}(t,\phi,E))$ and $(P(\tau,\phi,E),
X(\tau,\phi,E))$ to systems \eqref{maup1} (a) and (b) respectively with initial data on $\Lambda^1$.
Due to general properties of Hamiltonian systems, we have
\begin{equation*}
\mathcal{H}(\mathcal{X}(t,\phi,E),\mathcal{P}(t,\phi,E),E)=0 \quad\text{and}\quad
|P(\tau,\phi,E)| \; C(X(\tau,\phi,E))=1
\end{equation*}
and because of Maupertuis -- Jacobi correspondence, trajectories
$(\mathcal{P}(t,\phi,E),\mathcal{X}(t,\phi,E))$ and $(P(\tau,\phi,E), X(\tau,\phi,E))$ coincide modulo
a reparametrization of time. Indeed one has
\begin{gather}\nonumber
\frac{d \mathcal{P}}{d t}=-\mathcal{H}_x(\mathcal{P},\mathcal{X})=-R(\mathcal{X})H_x(\mathcal{P},\mathcal{X})=
  R(\mathcal{X})\frac{d P}{d \tau}
\\
\frac{d \mathcal{X}}{d t}=\mathcal{H}_p(\mathcal{P},\mathcal{X})=R(\mathcal{X})H_p(\mathcal{P},\mathcal{X})=
  R(\mathcal{X})\frac{d X}{d \tau}\label{maup2}
\end{gather}
where
$$R(x)=\lim_{z \to 1/C(x,E)}\quad  \frac{F(x,z)-E}{zC(x,E)-1}=
z \frac{\pa F}{\pa z}\bigl(x,z\bigr)|_{z=1/C(x,E)}$$
Changing time $t$ by time  $\tau=\tau(t,\phi,E),$ by using the equation
\begin{equation}\label{time}
   \frac{ d \tau}{dt}= R(\mathcal{X}(t,\phi,E)),\quad \tau|_{t=0}=0,
\end{equation}
we get (inverting the equation $\tau=\tau(t,\phi,E)$ we obtain $t=t(\tau,\phi,E)$)
\begin{gather}
\big(\mathcal{P}(t,\phi,E),\mathcal{X}(t,\phi,E)\big) =
\big(P(\tau,\phi,E), X(\tau,\phi,E)\big)\big|_{\tau=\tau(t,\phi,E)} \quad \Longleftrightarrow \quad
\\ \nonumber
\big(P(\tau,\phi,E), X(\tau,\phi,E)\big) =
\big(\mathcal{P}(t,\phi,E),\mathcal{X}(t,\phi,E)\big)\big|_{t=t(\tau,\phi,E)}.
\end{gather}
In $T^*\mathbb{R}^2$ the solutions $(P(\tau,\phi,E),
X(\tau,\phi,E))$ and $(\mathcal{P}(t,\phi,E),\mathcal{X}(t,\phi,E))$
define the phase flows (that we assume to be defined for all time)
\begin{align}\nonumber
\Lambda^2=
&\bigcup_{t\in\mathbb{R}}g^t_{\mathcal{H}}\Lambda^1=\{(p,x)=(\mathcal{P}(t,\phi,E), \mathcal{X}(t,\phi,E)),\,\phi\in {Q},\, t\in\mathbb{R}\} =
\\ \label{Lambda}
&\bigcup_{t\in\mathbb{R}}g^t_{H}\Lambda^1=\{(p,x)=(P(\tau,\phi,E), X(\tau,\phi,E)),\,\phi\in{\cal Q},\, \tau\in\mathbb{R}\}
\end{align}
In particular $\Lambda^2$ is invariant under $g^t_{\mathcal{H}}$ and $g^t_{H}$. Note that we could replace $\mathbb{R}^2$ by an open domain
$\Omega\subset\mathbb{R}^2$ and consider instead the maximal classical trajectories $g^t_{H}(\rho)$, $\rho\in\Lambda^1$ and $t\in(T_-(\rho),T_+(\rho))$,
and similarly for $g^t_{\mathcal{H}}$.
The parameters $t$ and $\tau$ are called {\it proper times}. Once $\Lambda^2$ is a smooth manifold, it becomes an embedded Lagrangian manifold,
and $(t,\phi)$ and $(\tau,\phi)$  are just two different coordinate systems on $\Lambda^2$.
It is convenient to relate objects belonging to either Hamiltonians, such as eikonals or half-densities. In particular:

\textbf{Lemma 1.}  The following properties hold:\\
\newline
1) The Jacobians of the transformation $(t,\phi)\mapsto(\tau,\phi)$ or its inverse verify
\begin{gather}\label{Jac1}
\det\frac{\pa (\tau(t,\phi,E),\phi)}{\pa(t,\phi)}=
\frac{d \tau}{dt}= R(\mathcal{X}(t,\phi,E)),
\\ \label{Jac2}
\det\frac{\pa (t(\tau,\phi,E),\phi)}{\pa(\tau,\phi)}=
\frac{d t}{d\tau}= 1/R(X(\tau,\phi,E))
\end{gather}
and the Jacobians $J=\det\frac{\pa X}{\pa(\tau,\phi)}$ (resp. $\mathcal J=\det\frac{\pa \mathcal X}{\pa (t,\phi)}$)
in coordinates $(\tau,\phi)$ (resp. $(t,\phi)$) are related by
\begin{equation}\label{Jac}
{\cal J}(t,\phi)=R(\mathcal{X}(\tau,\phi,E)) J(\tau,\phi)\bigg|_{\tau=\tau(t)}.
\end{equation}
\newline
2) The action function (eikonal) on $\Lambda^2$ is
\begin{gather}\nonumber
  s(t,\phi)\equiv\int_{(0,0)}^{(t,\phi)} \mathcal{P}(t,\phi,E)\,d\mathcal{X}(t,\phi,E)=
s_0(\phi)+\tau\\ \label{Eik}
s_0(\phi)=\int_{0}^{\phi} P^0(\phi,E)\,d X^0(\phi,E)
\end{gather}
3) The Jacobians $J,{\cal J}$ satisfy to the relations:
\begin{equation}\label{Jac0}
|J| =C(X(\tau,\phi))|X_\phi|, \quad|\mathcal J| = R\big(\mathcal X(t,\phi)\big) C\big(\mathcal X(t,\phi)\big) |X_\phi|.
\end{equation}

\textbf{Proof.} The first equalities \eqref{Jac1}, \eqref{Jac2}, \eqref{Jac}
hold since $\frac{\pa \phi}{\pa \tau}=\frac{\pa \phi}{\pa t}=0$.
The proof of \eqref{Eik} follows from chain of equalities
\begin{gather*}\nonumber
   s(t,\phi)= \int_{(0,0)}^{(t,\phi)} \mathcal{P}(t',\phi,E)\,d\mathcal{X}(t',\phi,E)=\int_{(0,0)}^{(\tau,\phi)} P(\tau',\phi,E)\,d X(\tau',\phi,E)=\\
\nonumber\int_{0}^{\phi} P^0(\phi,E)\,d X^0(\phi,E)+\int_{0}^{\tau} P(\tau',\phi,E)\,\frac{ d X}{d \tau'}(\tau',\phi,E) d\tau'= \\
\int_{0}^{\phi} P^0(\phi,E)\,d X^0(\phi,E)+\int_{0}^{\tau} |P(\tau',\phi,E)|\; C(X(\tau',\phi,E),E) d\tau'=s_0(\phi)+\tau
\end{gather*}
The assertion 3) is proved in \cite{DobrTirShaf,DMNT},
using that ${H}(x,p,E)$ is a Finsler symbol. $\Box$

The pair $(\widetilde\tau=s_0(\phi)+\tau, \phi)$ are called {\it eikonal coordinates} on $\Lambda^2$ (see \cite{DMNT}).
There are two important examples of curves $\Lambda^1$ in applications:
$\Lambda^1_s=\{p_1=0, p_2=k, x_1=\phi, x_2=a, \phi\in \mathbb{R}\}$ 
which appears in scattering problems and $\Lambda^1_G=\{p_1=b\cos \phi, p_2=b\sin \phi, x_1=a_1, x_2=a_2, \phi\in \mathbb{R}/2\pi\mathbb Z\}$
which appears in problems about the Green functions.
Easy to check that in these cases $s_0(\phi)=0$ and $\widetilde\tau=\tau$.

\section{Relationship with Maslov canonical operator}

We endow the Lagrangian manifold $\Lambda^2$ with
the measure $d\mu = dt\wedge d\phi$; let $A(t,\phi)$ be a smooth function on $\Lambda^2$ and
\begin{equation}\label{KO}\psi=K_{\Lambda^2}^h A(t,\phi).
\end{equation}
where $K_{\Lambda^2}^h$ is Maslov canonical operator.
We want to pass in $K_{\Lambda^2}^h A(t,\phi)$  from coordinates $(t,\phi)$ to eikonal-coordinates $(\tau,\phi)$ preserving the measure $d\mu$.

\textbf{Theorem.} The following equalities hold:
\begin{gather}\nonumber
\psi = K_{\Lambda^2}^h \Bigl[A(t(\tau,\phi),\phi)\Big/\sqrt{\det\frac{\pa(\tau,\phi)}{\pa (t,\phi)}}\Bigr]=
\\
K_{\Lambda^2}^h \Bigl[\frac{A(t(\tau,\phi),\phi)}{\sqrt{R(X(\tau,\phi))}}\Bigr]=
\frac{1}{\sqrt{R(x)}}K_{\Lambda^2}^h \Bigl[A(t(\tau,\phi),\phi)\Bigr](1+{\cal O}(h)).
\label{KO2}
\end{gather}
\textbf{Proof.}
It follows easily from \eqref{Jac1} and the commutation formula between the Pseudodifferential operator
$\hat Q = Q(x,hD_x)$ and Maslov canonical operator \cite{MasFed,Maslov}:
$\hat Q K_{\Lambda^2}^h \big[A(t,\phi)\big] =
K_{\Lambda^2}^h \big[Q(x,p)|_{\Lambda^2} \; A(t,\phi)\big] (1+{\cal O}(h))$. $\Box$

Recall that the the canonical operator has different representations in the neighborhood of regular points
(where $J = \det \frac{\pa X}{\pa(\tau,\phi)}\neq 0$)
and in the neighborhood of singular (focal) points
(where $J =\det \frac{\pa X}{\pa(\tau,\phi)} = 0$).
According to \eqref{Jac0} the point $(P(\tau,\phi),X(\tau,\phi))\in \Lambda ^2$ is singular (focal) if $X_\phi(\tau,\phi)=0$.
It was proved in \cite{DMNT} that under existence of the eikonal coordinates,
$\det(P,P_\phi)\neq 0$ in the neighborhood of the focal points.
Here $(P, P_\phi)$ is the $2\times 2$ matrix constructed from vector columns $P$ and $P_\phi$.
Thus the Lagrangian manifold could be covered by regular charts $\Omega_j^{\reg}$ with
$X_\phi(\tau,\phi)\neq 0$ and singular charts $\Omega_j^{\sing}$ with $\det (P,P_\phi)\neq 0$.
Let $\{\mathbf{e}_j(\tau,\phi)\}$ be a (finite) partition of unity subordinated to the charts $\Omega_j^{\reg}$, $\Omega_j^{\sing}$. 
Then due to Lemma 1 the contribution of a regular chart to the canonical operator is
\begin{equation}\label{reg}
\psi_j = \frac{e^{-i\frac{\pi}{2} \mathbf{m}_j}}{\sqrt{R(x)C(x)|X_\phi|}} 
e^{i\frac{\tau}{h}}A(\tau,\phi)\mathbf{e}_j(\tau,\phi)\Big|_{(\tau,\phi) = (\tau_j(x),\phi_j(x))}
\end{equation}
where $(\tau_j(x),\phi_j(x))$  is the solution to the (vector) equation $X(\tau,\phi)=x$ in the chart  
$\Omega_j^{\reg}$ and $\mathbf{m}_j$ is the Maslov index of $\Omega_j^{\reg}$ (see below).
The contribution of a singular chart is \cite{DMNT}
\begin{equation}\label{sing}
  \psi_j=\frac{e^{-i\frac{\pi}{2} \mathbf{m}^s_j}e^{i\frac{\pi}{4}}}{\sqrt{h R(x)}} \int_{\mathbb{R}}
  e^{i\frac{\tau}{h}}\sqrt{|\det(P,P_\phi)|}A(\tau,\phi)\mathbf{e}_j(\tau,\phi)\Big|_{\tau=\tau_j(x,\phi)}d\phi
\end{equation}
where $\tau_j(x,\phi)$  is the solution to the scalar equation $\langle P(\tau,\phi), x - X(\tau,\phi)\rangle=0$ in the chart  
$\Omega_j^{\sing}$ and $\mathbf{m}_j^s$ is the Maslov index of $\Omega_j^{\sing}$.

According to \cite{DMNT} Maslov index $\mathbf{m}_j$ coincides with Morse index of the trajectory starting 
from the point $(P,X)$ with coordinates $(\tau=0^+,\phi)$ and coming to the point $(P(\tau,\phi),X(\tau,\phi))\in\Omega_j^{\reg}$:
it equals to a number of zeroes of Jacobian $J = \det \frac{\pa X}{\pa(\tau',\phi)}$
(or the function $X_\phi(\tau',\phi)$)
when $\tau'$ runs from $0^+$ to $\tau$.
To find the index $\mathbf{m}_j^s$ of a singular chart $\Omega_j^{\sing}$
one need to take an arbitrary regular point $(P(\tau,\phi),X(\tau,\phi))\in \Omega_j^{\sing}$
and compare the signs of $J = \det \frac{\pa X}{\pa(\tau',\phi)} $ and $\det(P,P_\phi)$.
Then $\mathbf{m}_j^s$ equals Morse index of $(P(\tau,\phi),X(\tau,\phi))$ if they coincide,
and Morse index plus 1 otherwise.
Finally to construct the canonical operator one should patch all $\psi_j$ together (see \cite{Maslov,MasFed}).
At last note that integral \eqref{sing} could be expressed in the form of Airy or Pearcey functions
(see explicit formulas in \cite{DMNT})
under the assumption that the certain subset of Lagrangian singularities
$\{(P(\tau,\phi), X(\tau,\phi))|_{X_\phi=0}\}$ are in the so-called general position (\cite{Arnold, MasFed}).

We consider the following example.
The Lagrangian manifold presented in Fig. 1 has 2 caustics (red lines).
Under the area in configuration space between edges of a caustic
the Lagrangian manifold is folded into 3 leaves.
So in this area 3 functions of the form \eqref{reg} are to be patched together.
Under the area ``outside caustics'' there is only one leave of the manifold,
equation $X(\tau,\phi) = x$ has a unique solution and the canonical operator takes the form of \eqref{reg}
with a single function.
In the vicinity of caustic edges canonical operator is a sum of a regular \eqref{reg} and singular \eqref{sing} parts.

\section{Examples}

Let us present several examples of application of the Maupertuis -- Jacobi correspondence
for the construction of Maslov canonical operator.
We do not discuss here further applications to Partial Differential Equations.

{\bf Example 1} (from the Schr\"odinger equation, see \cite{Vainb,Kuch}).
Let $U(x)$ be a smooth bounded function, $U(x)<E$. Consider the classical Hamiltonian $\mathcal{H}(x,p) = F(x,|p|) = \frac{p^2}{2}+U(x)$.
Then
\begin{equation}\label{Schr}
C(x,E) = \frac{1}{\sqrt{2(E-U(x))}}, \quad R(x) = z^2\Big|_{z={1/C(x)}}=2(E-U(x))
\end{equation}
and
\begin{equation}\label{KO3}
\psi(x) = \frac{1}{\sqrt{2(E-U(x))}} \;
K_{\Lambda^2}^h \Bigl[A(t(\tau,\phi),\phi)\Bigr].
\end{equation}

{\bf Example 2} (from the two-dimensional Dirac equation for graphene, \cite{Katsnelson}).
Let $U(x), m(x)$ be smooth bounded functions. Consider the effective Hamiltonians
$\mathcal{H}^\pm(x,p) = F(x,|p|) = U(x)\pm\sqrt{p^2+m(x)^2}$.
Then
\begin{equation}\label{graph1}
C(x,E)=\frac{1}{\sqrt{(E-U)^2-m^2}}, \quad
R=\pm \frac{z^2}{\sqrt{z^2+m(x)^2}}\Big|_{z={1/C}}=\frac{(E-U(x))^2-m^2(x)}{E-U(x)},
\end{equation}
and
\begin{equation}\label{graph2} \psi= \frac{\sqrt{E-U(x)}}{\sqrt{(E-U(x))^2-m(x)^2}} \;
K_{\Lambda^2}^h \Bigl[A(t(\tau,\phi),\phi)\Bigr].\end{equation}

{\bf Example 3} (from the water waves theory, \cite{DR2,Dobr,DobrZh}).
Let $D(x)>0$ be the smooth function, 
representing the depth of a basin, and consider the effective Hamiltonian $\mathcal{H}(x,p) = F(x,|p|) = \sqrt{|p| \; \tanh(|p| D(x))}-E$.
It is easy to see that there exists a unique smooth positive solution $y=Y(\mathcal{E}(x))$ to the equation
$\sqrt{y \tanh(y)}=\mathcal{E}(x)=\sqrt{D(x)}E$ and
\begin{gather}\nonumber
C(x,E)=\frac{D(x)}{y\big(E\sqrt{ D(x)}\;\big)} , \quad
R = z \frac{D(x) z/\cosh^2(zD(x)) + \tanh(zD(x))}
{2 \sqrt{z \tanh(z D(x))}}\Big|_{z=\frac{1}{C}}=\\
\frac{(y^2-y^2\tanh^2(y)) + y\tanh(y))}{2\sqrt{D} \sqrt{y \tanh(y)}}\Big|_{y=Y\big(\sqrt{D(x)}E\big)}=
\frac{y^2 - D(x)^2 E^4 + D(x) E^2}{2D(x) E }\Big|_{y=Y\big(\sqrt{D(x)}E\big)}\label{ww}
\end{gather}

The Hamiltonian system with the Hamiltonian $H(x,p,E) = C(x,E)|p|$ has the form
\begin{gather}
\frac{dp}{d\tau} = - |p| \frac{\pa}{\pa x} \bigg(\frac{D(x)}{Y\big(E\sqrt{ D(x)}\;\big)} \bigg),
\qquad
\frac{dx}{d\tau} = \frac p {|p|}  \frac{D(x)}{Y\big(E\sqrt{ D(x)}\;\big)}.
\end{gather}
It contains function $Y(\mathcal E(x))$ and its derivative $Y'(\mathcal E(x))$
which makes difficult to finding its inverse.
Let us show how to rewrite this system in a form without the function $Y(\mathcal E(x))$.

Along the trajectories $(P,X)$ the following equalities hold:
\begin{equation} \label{eq_y}
H(X,P,E) = 0  \quad \Leftrightarrow \quad
C(X,E)|P| = 1 \quad \Rightarrow \quad
Y(E\sqrt{ D(X)}) = \frac {D(X)}{C(X,E)} = D(X)|P|.
\end{equation}
Then differentiating the equation for $Y(\mathcal E(x))$ we have
\begin{gather}\label{ham1}
Y' \big(\tanh Y + Y (1 - \tanh^2 Y)\big) = 2\sqrt{Y \tanh Y}.
\end{gather}
This gives for the solution $(P,X)$
\begin{gather}\label{eq_dy}
Y'(E \sqrt{D(X)}) = \frac{2 Y \mathcal E}{Y^2 + \mathcal E^2 - \mathcal E^4}\bigg|_{(P,X)} =
\frac{2 |P| \sqrt{D(X)} E}{D(X) |P|^2 + E^2 - D(X) E^4}.
\end{gather}
Inserting these equalities \eqref{eq_y} and \eqref{eq_dy} into the Hamiltonian system we finally have
\begin{gather}\label{HamSys_Ex3}
\frac{dp}{d\tau} = -\frac{p^2-E^4}{D(x) p^2 + E^2 - D(x) E^4}\cdot \frac{\pa D(x)}{\pa x},
\qquad
\frac{dx}{d\tau} = \frac p {p^2}
\end{gather}

To write out the canonical operator we also insert $Y|_{(P,X)}$ into the expression \eqref{ww} for $R$:
\begin{equation}\label{R_Ex3}
R|_{(P,X)} \equiv R(X,P,E) = \frac{D(X) P^2 - D(X) E^4 +  E^2}{2 E}
\end{equation}
Taking into account the last expression for $R$ and that $C(X) = 1/|P|$
we obtain the formula \eqref{reg} for canonical operator in a regular point in this case
\begin{equation}\label{sol_reg_Ex3}
\psi_j(x) = \frac{e^{i\frac{\tau}{h}} e^{-i\frac{\pi}{2} \mathbf{m}_j}}{\sqrt{|X_\phi(\tau,\phi)|}}
\sqrt{\frac{2 E \;|P(\tau,\phi)|}{D\big(X(\tau,\phi)\big) P^2 - D\big(X(\tau,\phi)\big) E^4 +  E^2}}
\;\; A(\tau,\phi)\mathbf{e}_j(\tau,\phi)\Big|_{(\tau,\phi) = (\tau_j(x),\phi_j(x))}
\end{equation}

As $R$ depends on $p$, it is not convenient to factor out $1/\sqrt{R(x,hD_x,E)}$ from the canonical operator. 
Near the focal point we write instead
\begin{equation}\label{sol_sing_Ex3}
\psi_j(x) =
e^{-i\frac{\pi}{2} \mathbf{m}^s_j}e^{i\frac{\pi}{4}}\frac{\sqrt{2E}}{\sqrt{h}}
\int_{\mathbb{R}}   \frac{\sqrt{|\det\big(P(\tau,\phi),P_\phi(\tau,\phi)\big)|} \;\;
                          e^{i\frac{\tau}{h}}A(\tau,\phi)\mathbf{e}_j(\tau,\phi)}
{\sqrt{D\big(X(\tau,\phi)\big) P^2(\tau,\phi) - D\big(X(\tau,\phi)\big) E^4 + E^2}}\Big|_{\tau=\tau_j(x,\phi)}d\phi
\end{equation}

{\bf Example 4} (from the water waves theory with surface tension, \cite{Dobr,DobrZh,DR2}).
We modify Hamiltonian in Example 3 according to
$\mathcal{H}(x,p) = F(x,|p|)-E = \sqrt{|p| \; \tanh(|p| D(x))(1+\mu(x)|p|^2)}-E$, $x\in\mathbb{R}^2$,
where $\mu(x)>0$ is a smooth function, representing the surface tension of the fluid. Let also $\nu(x)=E(\mu(x))^{1/4}$,
${\cal E}(x)=E(D(x))^{1/2}$. The relation $\mathcal{H}(x,p)=0$ can be rewritten in a functional form as
$f(y,{\cal E},\nu)=0$, where $f(y,{\cal E},\nu)=y\tanh(y)-{\cal E}^2\bigl(1+y^2\frac{\nu^4}{{\cal E}^4}\bigr)^{-1}$ is smooth
on $\mathbb{R}_+^3$, and because $\frac{\partial f}{\partial y}(y,{\cal E},\nu)>0$,
the implicit functions theorem gives $y=Y({\cal E},\nu)$ where $Y$ is smooth in $({\cal E},\nu)\in{\bf R}_+^2$.
Since $\mu$ and $D$ are smooth functions, it follows also from the implicit function theorem in Fr\'echet space $C^\infty(\mathbb{R}^2)$
that $y=Y({\cal E},\nu)\in C^\infty(\mathbb{R}^2)$.
As above, the equations of motion with Hamiltonian $C(x,E)|p|$ have the form
\begin{gather}
\frac{dp}{d\tau} = - |p| \frac{\pa}{\pa x} \bigg(\frac{D(x)}{Y\big(E\sqrt{ D(x)},E{\mu(x)}^{1/4}\;\big)} \bigg),
\qquad
\frac{dx}{d\tau} = \frac p {|p|}  \frac{D(x)}{Y\big(E\sqrt{ D(x)},E{\mu(x)}^{1/4}\;\big)}.
\end{gather}
As above they simplify on $H(x,p,E)=C(x,E)|p|-1=0$ to
\begin{gather}\label{HamSys_Ex4}
\frac{dp}{d\tau}=
-|p| \bigg(\frac 1{y} \frac{\pa D}{\pa x}  - \frac {D(x)}{y^2} \frac{d Y}{d x}\bigg)
\bigg|_{y = D(x)|p|}, \qquad
\frac{dx}{d\tau} = \frac p {p^2}
\end{gather}
Here $dY/dx$ is found by differentiating the equation $y = Y({\cal E}(x),\nu(x))$:
\begin{gather*}
\frac{d Y}{d x}=-\bigl(\frac{\pa f}{\pa y}\bigr)^{-1}\bigl(\frac{\pa f}{\pa {\cal E}}\frac{\pa {\cal E}}{\pa x}+
\frac{\pa f}{\pa \nu}\frac{\pa \nu}{\pa x}\bigr),
\\
\frac{\pa f}{\pa y}=\tanh(y)+y(1-\tanh^2(y))+2y{\cal E}^{-2}\nu^4\bigl(1+y^2\frac{\nu^4}{{\cal E}^4}\bigr)^{-2}>0.
\end{gather*}
Substituting this derivatives into \eqref{HamSys_Ex4} with $y(E\sqrt{D(X)}) = |P|D(X)$ we get a system similar to \eqref{HamSys_Ex3}, 
and can we obtain also an expression for $R$ as in \eqref{R_Ex3}.
So we are able to get a representation of Maslov canonical operator as in \eqref{sol_reg_Ex3} and \eqref{sol_sing_Ex3}.

\begin{acknowledgments}

{This work was supported by RFBR grant N~14-01-00521 and by the Russian Federation President Programm MK-1017.2013.1.
S.~Yu.~D. and D.~S.~M. are grateful to the staff of Centre de Physique Th\'eorique and Universit\'e du Sud Toulon-Var for support and kind hospitality.
Authors are grateful to V.~E.~Nazaikinskii for fruitful discussions.}

\end{acknowledgments}

\newpage
\begin{figure}\label{Fig_LagMan}
\setcaptionmargin{5mm}
\onelinecaptionsfalse
\includegraphics{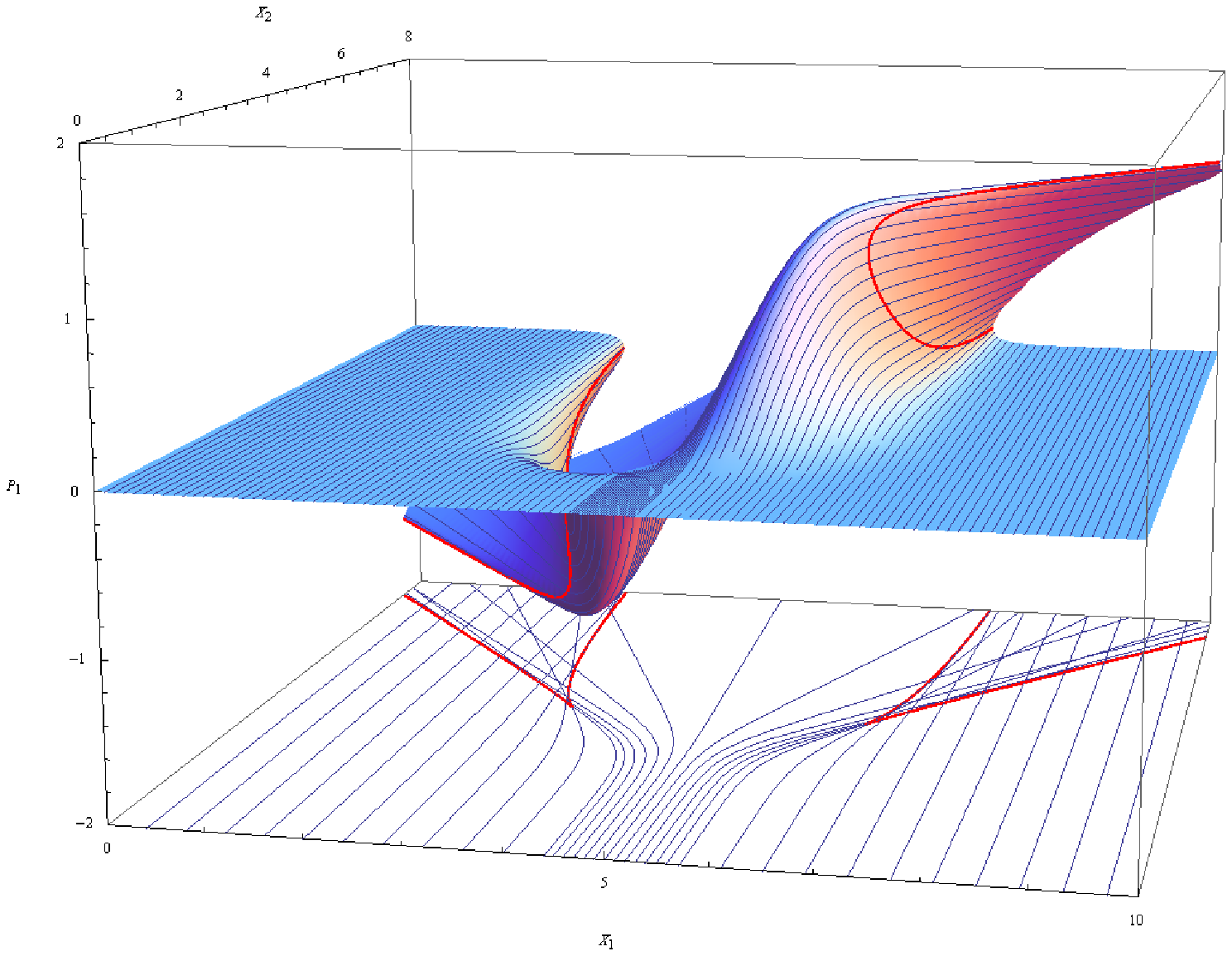}
\captionstyle{normal}
\caption{Lagrangian manifold, characteristics (blue lines) and caustic (red lines):
in the phase space (in coordinates $(x_1,x_2,p_1) \subset \mathbb R^4_{x,p}$)
and in projection to the configuration space $\mathbb R^2_x$.
The Lagrangian manifold $\Lambda^2 = \bigcup g^t_H \Lambda^1$ corresponds to a scattering problem with initial curve
$\Lambda^1 = \{p = (0,2), x=(\phi,0), \phi\in\mathbb R \}$,
the Hamiltonian is $H(x,p) = |p|/(E-U(x))$,
where $E = 2, U(x) = {\bf e}(x) e^{- (x_1 - 5)^2 - (x_2 - 3)^2} $ and ${\bf e}(x)$ -- is a cut-off function
${\bf e}(x) = 0, x_2\leq0, \; {\bf e}(x) = 1, x_2\geq1$.}
\end{figure}


\begin{thebibliography}{99}



\bibitem{Arnold}
Arnold~V.~I. Mathematical Methods of Classical Mechanics.
Berlin: Springer-Verlag, 1978.

\bibitem{Ab}
Abraham~R., Marsden~J.~E.  Foundations of Mechanics.
American Math. Soc., 1978.

\bibitem{Tsi}
A.~V.~Tsiganov,
The Maupertuis Principle and Canonical Transformations of the Extended Phase Space.
J. of Nonlinear Math. Physics, {\bf 8}, p.157–182 (2001).

\bibitem{Taylor}
Taylor~M.
Finsler structures and wave propagation,
in {\it Isakov~V.} (ed.) Sobolev spaces in Mathematics III. Int. Math.
Series, Springer, 2009.

\bibitem{DR1}
S.~Yu.~Dobrokhotov, M.~Rouleux,
The Semiclassical Maupertuis -- Jacobi Correspondence and  Applications to Linear Water Wave Theory ,
Mathematical Notes, {\bf 87}, 3, 430 (2010)

\bibitem{DR2}
S.~Yu.~Dobrokhotov, M.~Rouleux,
The semi-classical Maupertuis-Jacoby correspondence for quasi Periodic Hamiltonian flows with applications to  linear water waves theory,
Asymptotic Analysis, {\bf 74}, 1-2, 33 (2011)

\bibitem{Katok}
A.~B.~Katok, Ergodic properties of degenerate integrable Hamiltonian systems,
Math. USSR-Izv. {\bf 7} 535 (1973)

\bibitem{DMNT}
S.~Yu.~Dobrokhotov, G.~Makrakis, V.~E.~Nazaikinskii, and T.~Ya.~Tudorovskii,
New formulas for Maslov canonical operator in a neighborhood of focal points and caustics in 2D semiclassical asymptotics,
Theor. and Math. Physics,  {\bf 177}, 3, 1579 (2013)

\bibitem{MasFed}
Maslov~V.~P., Fedoriuk~M.~V.
Semi-classical approximation in Quantum Mechanics.
Reidel, 1981.

\bibitem{Maslov}
Maslov~V.~P.
Th{\'e}orie des Perturbations et M{\'e}thod Asymptotiques.
Dunod, Paris, 1972.

\bibitem{DobrTirShaf}
S.~Yu.~Dobrokhotov, A.~I.~Shafarevich, B.~Tirozzi,
Localized Wave and Vortical Solutions to Linear Hyperbolic Systems and Their Application to the Linear Shallow Water Equations,
Russ. J. Math. Phys., {\bf 15}, 2, 192 (2008)

\bibitem {Vainb}
Vainberg~B.~R.
Asymptotic methods in equation of mathematical physics.
New York, Gordon \& Breach Science Publishers, 1989.

\bibitem{Kuch}
V.~V.~Kucherenko,
Quasiclassical asymptotics of a point-source function for the stationary Schrodinger equation,
{Teor. Math. Phys.}, {\bf 1}, 3, 294 (1969)

\bibitem{Katsnelson}
Katsnelson~M.
Graphene: Carbon in Two Dimensions.
Cambridge University Press, 2012.


\bibitem{Dobr}
S.~Yu.~ Dobrokhotov,
Maslov's methods in linearized theory of gravitational waves on the liquid surface,
Sov. Phys. Doklady, {\bf 28}, 229 (1983)


\bibitem{DobrZh}
S.~ Dobrokhotov, P.~Zhevandrov,
Asymptotic expansions and the Maslov canonical operator in the linear theory of~water waves. I.~Main constructions and equations for surface gravity waves.
{Russ. J. Math. Phys.}, {\bf 10} (2003)

\end{thebibliography}
\end{document}